\documentclass[twocolumn,showkeys,showpacs,amsmath,amssymb,superscriptaddress,preprintnumbers,nofootinbib]{revtex4}
\usepackage{amssymb}
\usepackage{amsmath}
\usepackage{graphicx}

\normalbaselines
\begin{document}

\title{Birefringence induced by pp-wave modes \\ in an electromagnetically active  dynamic aether}

\author{Timur Yu. Alpin}
\email{Timur.Alpin@kpfu.ru}
\author{Alexander B. Balakin}
\email{Alexander.Balakin@kpfu.ru}
\affiliation{Department of General Relativity and
Gravitation, Institute of Physics,Kazan Federal University, Kremlevskaya street 18, Kazan, 420008,
Russia}

\begin{abstract}
In the framework of the Einstein-Maxwell-aether theory we study the birefringence effect, which can occur in the pp-wave symmetric dynamic aether.
The dynamic aether is considered to be latently birefringent quasi-medium, which displays this hidden property if and only if the aether  motion is non-uniform, i.e., when the aether flow is characterized by the non-vanishing expansion, shear, vorticity or acceleration. In accordance with the dynamo-optical scheme of description of the interaction between electromagnetic waves and the dynamic aether, we shall model the susceptibility tensors by the terms linear in the covariant derivative of the aether velocity four-vector. When the pp-wave modes appear in the dynamic aether, we deal with a gravitationally induced degeneracy removal with respect to hidden susceptibility parameters. As a consequence, the phase velocities of electromagnetic waves possessing orthogonal polarizations do not coincide, thus displaying the birefringence effect. Two electromagnetic field configurations are studied in detail: longitudinal and transversal with respect to the aether pp-wave front. For both cases the solutions are found, which reveal anomalies in the electromagnetic response on the action of the pp-wave aether mode.
\end{abstract}



\maketitle

\section{Introduction}

The effect of birefringence is well documented in the electrodynamics of continuous media \cite{LL,Eringen,Hehlbook,Botcher1,Botcher2}. This effect reveals itself, in particular, when the electromagnetic waves possessing two orthogonal polarizations are forced to move with different phase velocities, thus being converted to the so-called ordinary and extraordinary waves. The medium behaves as the birefringent one, when the electric and magnetic susceptibility  tensors of the medium are anisotropic, i.e., when these tensors possess non-coinciding eigen-values (the medium is called bi-axial, if all three eigen-values are different, and uni-axial, when only two of them coincide). The birefringent property of the medium can be the intrinsic one (e.g., in the spatially anisotropic crystals \cite{Sirotin}, in moving uni-axial media \cite{BZ}), or can be induced by external influences (e.g., by an external electric field \cite{Botcher1,Botcher2}, magnetic field \cite{M1,M2},  stresses, anisotropic heating, etc., \cite{Sirotin}). When we deal with electromagnetic waves propagating under the influence of the gravitational field, various versions of the gravitation theory predict different results. For instance, the pre-metric axiomatic theory guarantees (see, e.g., \cite{No}) that there is no intrinsic birefringence. Similarly, the minimal Einstein version of the theory of gravity excludes birefringence. However, in the framework of the modified theories of gravity  the effect of birefringence was predicted by many authors. For instance, the nonminimal Einstein--Maxwell theory admits the birefringence effect since the coupling of photons to the curvature makes the nonminimal susceptibility tensor anisotropic (see, e.g., \cite{DH,LM,B97}). Violation of the Lorentz invariance of the model \cite{L1,L2,L3,L4,L5}, a torsion nonminimally coupled to photons \cite{Torsion}, interactions with strings \cite{string},  also  can be the origin of the birefringence effect. These predictions have attracted the attention to the problem of cosmic birefringence and its observations \cite{CB1,CB2,CB3,CB4,CB5}.

Our goal is to study the birefringence induced by the dynamic aether. We assume that when the motion of the aether is uniform, the aether is not birefringent, i.e.,  the effect we search for is hidden. In other words, when  the motion of the aether is uniform, the test electromagnetic waves do not display the dependence of phase on the polarization; however, when the aether flow is characterized by non-vanishing acceleration, shear, rotation or expansion, we deal with the so-called degeneracy removal with respect to the hidden parameters in analogy with effects described in \cite{deg1}. The idea of mathematical description of this degeneracy removal was disclosed in \cite{LL}; there the corresponding term {\it dynamo-optical phenomena} was introduced. In order to describe this effect the authors of \cite{LL} have introduced the terms with derivatives of the medium velocity into the permittivity tensors, thus rendering these tensors spatially anisotropic.

Our consideration is based on the Einstein-aether theory \cite{J1,J2,J3,J4,J5,J6,J7,J8,J9,J10} and its extension, the Einstein--Maxwell-aether theory \cite{BL2014}. The macroscopic velocity four-vector $U^i$ appears in the Einstein-aether and Einstein--Maxwell-aether theories as a dynamic time-like vector field normalized by unity ($g_{ik}U^iU^k=1$). The covariant derivative $\nabla_i U_k$ enters the basic Lagrangian of the Einstein-aether theory \cite{J1}, and it appears in the interaction terms in the Einstein--Maxwell-aether theory \cite{BL2014}. In this sense, we can indicate our approach as an {\it extension} of the idea of dynamo-optical interactions, fulfilled in the framework of the Einstein-aether theory.
The  Einstein-aether and the Einstein-Maxwell-aether theories realize the idea
of a preferred frame of reference \cite{CW,N1,N2} associated with a world-line congruence for which the corresponding time-like velocity four-vector $U^i$ is the tangent vector. In this sense they are characterized by a violation of Lorentz invariance (see, e.g., \cite{L3}).
There is also an alternative approach to introduce dynamo-optical interactions, which is based on the analysis of the time-like unit {\it eigen four-vector} of the stress-energy tensor of the cosmic substratum (the vacuum, the aether, the dark fluid and so on) (see, e.g., \cite{BDehnen,BDolbilova,BAlpin}). Such velocity field appears algebraically as an intrinsic vectorial quantity; the velocity field which we consider now is related to the additional dynamic vector field.

In this paper we consider the birefringence effect, which is dynamo-optically induced by the aether pp-wave modes. What does this mean? First, we consider the pp-wave background formed by the gravitationally self-interacting aether and fix the constraints on the Jacobson coupling parameters, which guarantee that the so-called pp-wave modes can exist in the dynamic aether.  Second, we study the propagation of test electromagnetic waves dynamo-optically coupled to the pp-wave symmetric background. The modeling of the susceptibility tensors of such potentially birefringent aether is based on the introduction of two coupling constants; the phenomenologically constructed susceptibility tensors describe some effective bi-axial quasi-medium. Then we analyze the master equations for the longitudinal and transversal electromagnetic field configurations, and prove that this aether behaves as a birefringent medium.

The paper is organized as follows. In Sect. 2 we consider the basic elements of the Einstein--Maxwell-aether theory, and describe the background state possessing the pp-wave symmetry and introduce a specific background state indicated as pp-wave aether mode. In Sect. 3 we study solutions for electromagnetic waves in the aether with excited pp-wave modes. In Sect. 4 we discuss the magnitudes of the birefringence effect, and demonstrate that anomalies can exist in the electromagnetic response on the action of the  pp-wave aether modes. Briefly our conclusions are presented in Sect. 5.

\section{The formalism}

\subsection{Action functional of the Einstein-aether theory}

The Einstein-aether theory \cite{J1,J2,J3,J4,J5,J6,J7,J8} uses the action functional
$$
{\cal S}_{({0})} = \int d^4 x \sqrt{{-}g} \
\frac{1}{2\kappa} \left[R{+} \lambda \left(g_{mn}U^m
U^n {-}1 \right) {+} \right.
$$
\begin{equation}
\left.
+K^{abmn} (\nabla_a U_m) (\nabla_b U_n)
\right]\,,
\label{1}
\end{equation}
which describes the interaction between the gravitational field and unit vector field $U^i$ attributed to the velocity of some hypothetic medium, the dynamic aether.
In the functional (\ref{1}), the quantity $g {=} {\rm det}(g_{ik})$ is the determinant of the metric; $R$ is the Ricci
scalar; $\kappa$ is the Einstein constant. The term $\lambda \left(g_{mn}U^m U^n {-}1 \right)$ ensures
that the $U^i$ is normalized to one; the function $\lambda$ is the Lagrange multiplier.
The term $K^{abmn} \
\nabla_a U_m \ \nabla_b U_n $ is quadratic in the covariant derivative
$\nabla_a U_m $ of the vector field $U^i$. The tensor $K^{abmn}$ is constructed
using the metric tensor $g^{ij}$ and the velocity four-vector $U^k$ only (see, e.g., \cite{J1}):
$$
K^{abmn} =
$$
\begin{equation}
C_1 g^{ab} g^{mn} {+} C_2 g^{am}g^{bn}
{+} C_3 g^{an}g^{bm} {+} C_4 U^{a} U^{b}g^{mn} \,.
\label{2}
\end{equation}
Here $C_1$, $C_2$, $C_3$ and $C_4$ are phenomenologically introduced coupling constants \cite{J1,J2,J3}.
In order to interpret the coupling constants $C_1$,$C_2$,$C_3$,$C_4$, one uses the standard decomposition  of the
tensor $\nabla_i U_k$ into the sum
\begin{equation}
\nabla_i U_k = U_i DU_k + \sigma_{ik} + \omega_{ik} +
\frac{1}{3} \Delta_{ik} \Theta \,. \label{act3}
\end{equation}
The acceleration four-vector $DU^{i}$,
symmetric trace-free shear tensor $\sigma_{ik}$,
anti-symmetric vorticity tensor $\omega_{ik}$, and
the expansion scalar $\Theta$ are given by the formulas
$$
DU_k \equiv U^m \nabla_m U_k \,,
$$
$$
\sigma_{ik}
\equiv \frac{1}{2}\Delta_i^m \Delta_k^n \left(\nabla_m U_n {+}
\nabla_n U_m \right) {-} \frac{1}{3}\Delta_{ik} \Theta  \,,
$$
$$
\omega_{ik} \equiv \frac{1}{2}\Delta_i^m \Delta_k^n \left(\nabla_m
U_n {-} \nabla_n U_m \right)  \,,
$$
\begin{equation}
 \Theta \equiv \nabla_m U^m  \,,
\quad \Delta^i_k = \delta^i_k - U^iU_k \,. \label{act4}
\end{equation}
In these terms the scalar
\begin{equation}
{\cal K} \equiv K^{abmn} (\nabla_a U_m) (\nabla_b U_n)
\label{act47}
\end{equation}
in the action functional (\ref{1}) can be rewritten as follows:
\begin{equation}
{\cal K} = {\cal C}_{D} DU_k DU^k {+}
{\cal C}_{\omega} \omega_{ik} \omega^{ik} {+} {\cal C}_{\sigma} \sigma_{ik} \sigma^{ik} {+}  \frac13  {\cal C}_{\Theta} \Theta^2
\,. \label{act5}
\end{equation}
Here we used the notations
$$
{\cal C}_{D} = C_1{+} C_4 \,, \quad {\cal C}_{\omega} = C_1{-} C_3 \,,
$$
\begin{equation}
{\cal C}_{\sigma} = C_1{+} C_3 \,, \quad {\cal C}_{\Theta} = C_1{+} 3C_2{+}C_3  \,.  \label{act950}
\end{equation}
As shown in \cite{J2}, the Einstein-aether theory admits waves of three types, which can be classified formally as scalar, vectorial, and tensorial; respectively, one can speak of waves with spin zero, spin one, and spin two. The parameters ${\cal C}_{D}$, ${\cal C}_{\omega}$, ${\cal C}_{\sigma}$ and ${\cal C}_{\Theta}$ are connected with velocities of the corresponding waves, denoted as $S_{(0)}$, $S_{(1)}$, and $S_{(2)}$. For weak waves on the Minkowski background these velocities of the waves are found to be (compare with \cite{J2})
\begin{equation}
S^2_{(0)} = \frac{({\cal C}_{\Theta}+ 2{\cal C}_{\sigma})(2-{\cal C}_{D})}{3{\cal C}_{D}(1-{\cal C}_{\sigma})(2+{\cal C}_{\Theta})} \,,
\label{act951}
\end{equation}
\begin{equation}
S^2_{(1)} = \frac{{\cal C}_{\sigma} + {\cal C}_{\omega}(1-{\cal C}_{\sigma})}{2{\cal C}_{D}(1-{\cal C}_{\sigma})}\,, \quad S^2_{(2)} = \frac{1}{(1-{\cal C}_{\sigma})}\,.
\label{act952}
\end{equation}
Our ansatz is that the tensorial mode propagates with the velocity coinciding with the speed of light in vacuum, i.e., $S_{(2)}=1$; this quantity also coincides with the standard velocity of propagation of the weak gravitational waves on the Minkowski background \cite{MTW}. According to (\ref{act952}) this means that ${\cal C}_{\sigma}{=}0$. Also, we assume that the $\{g,U \}$ model is pure vectorial-tensorial, and the scalar modes cannot propagate at all, $S_{(0)}{=}0$; then according to (\ref{act951}) we obtain ${\cal C}_{\Theta} = 0$. The velocity $S_{(1)}$ is free of restrictions; now we obtain $S_{(1)} = \sqrt{\frac{{\cal C}_{\omega}}{2{\cal C}_{D}}}$, and the coupling constants ${\cal C}_{\omega}$ and ${\cal C}_{D}$ are assumed to be of the same signs. Below we will show that these phenomenological motives lead to the same result as the strict definition of the pp-wave aether modes.

\subsection{Master equations describing the background state}

\subsubsection{Equations for the unit dynamic vector field}

The aether dynamic equations are known to be found by varying the action (\ref{1}) with
respect to the Lagrange multiplier $\lambda$ and to the unit vector field $U^i$.
The variation with respect to $\lambda$ gives the equation
\begin{equation}
g_{mn}U^m U^n = 1 \,,
\label{21}
\end{equation}
which is the normalization condition of the time-like
vector field $U^k$. Variation of the functional (\ref{1}) with respect to
$U^i$ yields that $U^i$ itself satisfies the standard balance equation
\begin{equation}
\nabla_a {\cal J}^{aj} =
 I^j + \lambda \ U^j  \,,
\label{00A1}
\end{equation}
where the auxiliary quantities ${\cal J}^{aj}$ and $I^j$ are defined as follows:
\begin{equation}
{\cal J}^{aj} \equiv  K^{abjn} (\nabla_b U_n)  \,, \quad I^j =  C_4 (DU_m)(\nabla^j U^m) \,.
\label{J2}
\end{equation}
The Lagrange multiplier $\lambda$ can be obtained by convolution of (\ref{00A1}) with $U_j$; it has the following form:
\begin{equation}
\lambda =  U_m \left[\nabla_a {\cal J}^{am}- I^m \right]  \,.  \label{0A309q}
\end{equation}
In more detail, using the constitutive tensor (\ref{2}) and the symbols $\Theta_{ik} \equiv \nabla_iU_k $, $\Theta \equiv \nabla_kU^k $, we obtain
$$
{\cal J}^{am} = C_1 \Theta^{am} {+} C_2 g^{am} \Theta
{+} C_3  \Theta^{ma} {+} C_4 U^{a} DU^{m}  \,,
$$
\begin{equation}
I^m =  C_4 (DU_n)\Theta^{mn} \,. \label{299}
\end{equation}

\subsubsection{Equations for the gravitational field}

The variation of the action (\ref{1}) with respect to the metric
$g^{ik}$ yields the gravitational field equations:
\begin{equation}
R_{ik} - \frac{1}{2} R \ g_{ik} = \lambda U_i U_k  + T^{({\rm U})}_{ik} \,. \label{0Ein1}
\end{equation}
The term $T^{({\rm U})}_{ik}$ describes
the stress-energy tensor associated with the self-gravitation
of the vector field $U^i$:
$$
T^{({\rm U})}_{ik} =
\frac12 g_{ik} {\cal J}^{am} \nabla_a U_m {+}
$$
$$
{+}\nabla^m \left[U_{(i}{\cal J}_{k)m}\right] {-}
\nabla^m \left[{\cal J}_{m(i}U_{k)} \right] {-}
\nabla_m \left[{\cal J}_{(ik)} U^m\right]+
$$
$$
+C_1\left[(\nabla_mU_i)(\nabla^m U_k) {-}
(\nabla_i U_m \nabla_k U^m) \right] {+}
$$
\begin{equation}
{+}C_4 (U^a \nabla_a U_i)(U^b \nabla_b U_k) \,.
\label{5Ein1}
\end{equation}
As usual, the symbol $p_{(i} q_{k)}{\equiv}\frac12 (p_iq_k{+}p_kq_i)$
denotes the procedure of symmetrization.
As will be shown below, the trace of this tensor,
$$
T^{({\rm U})} =
2 [C_1 \Theta^{am}\Theta_{am} {+} C_3  \Theta^{ma}\Theta_{am} ]  {+}
$$
$$
{+}\left(C_1 {+} C_4 {-}C_3 \right)\nabla_m DU^{m} {-}
\left(C_1+4C_2+C_3 \right)D\Theta -
$$
\begin{equation}
-
\left(C_1+2C_2+C_3 \right)\Theta^2
{+}3 C_4 DU^k DU_k
\label{trace}
\end{equation}
has to be equal to zero for the model with the pp-wave symmetry.

\subsection{Master equations reduced to the case with pp-wave symmetry}

\subsubsection{Metric and Killing's vectors}

We consider space-times, which possess the ${\rm G}_5$ group of isometries \cite{Exact}, with five Killing vectors $\{\xi^i_{(1)}, \xi^i_{(2)}, \xi^i_{(3)}, \xi^i_{(4)}, \xi^i_{(5)} \}$, three of which,  $\{\xi^i_{(1)}, \xi^i_{(2)}, \xi^i_{(3)} \}$, form the  Abelian subgroup ${\rm G}_3$, and the first of them, $\xi^i_{(1)}$, is the null covariantly constant four-vector.
Mathematically, this means that, first, the Lie derivative of the metric is equal to zero, $\pounds_{\xi^l_{({\rm a})}} g_{ik} =0$ (${\rm a} = 1,2,3,4,5$); second, $g_{ik}\xi^i_{(1)}\xi^k_{(1)}=0$, third, $\nabla_k \xi^i_{(1)}=0$.

The geometrical properties of space-times with this so-called pp-wave symmetry are well documented (see, e.g. \cite{Exact}). We use the metric in the TT-gauge
\begin{equation}
ds^2 = 2 du dv - L^2 \left(e^{2\beta} {dx^2}^2 +
e^{-2\beta} {dx^3}^2
\right) \,,
\label{PW1}
\end{equation}
which describes plane gravitational waves in the standard theory of gravity, and we choose for simplicity the wave with the first polarization (see, e.g., \cite{MTW} for details).
Here $u{=}\frac{1}{\sqrt2}(ct{-}x^1)$ and $v{=}\frac{1}{\sqrt2}(ct{+}x^1)$
are the retarded and advanced times, respectively.
The functions $L(u)$ and $\beta(u)$ are assumed to depend on the retarded time $u$ only, and to satisfy the conditions $L(0){=}1$, $L^{\prime}(0){=}0$, $\beta(0){=0}$, on the initial front plane $u{=}0$. The five Killing vectors in this representation are known to be of the form
$$
\xi^i_{(1)} = \delta^i_v \,, \quad  \xi^i_{(2)} = \delta^i_2 \,, \quad \xi^i_{(3)} = \delta^i_3 \,,
$$
\begin{equation}
\xi^i_{(4)} = x^2 \delta^i_v {+} G(u) \delta^i_2 \,, \quad \xi^i_{(5)} = x^3 \delta^i_v {+} F(u) \delta^i_3 \,,
\label{K1}
\end{equation}
where
$$
G(u)= \int_0^u du' L^{-2}(u')e^{-2\beta(u')} \,,
$$
\begin{equation}
F(u)= \int_0^u du' L^{-2}(u')e^{2\beta(u')} \,.
\label{K9}
\end{equation}
The four-vectors $\{\xi^i_{(1)}, \xi^i_{(2)}, \xi^i_{(3)} \}$ forming the Abelian subgroup ${\rm G}_3$ are orthogonal one to another.

\subsubsection{Ansatz about inheritance of the pp-wave symmetry}

We assume that the vector field $U^i$ inherits the pp-wave symmetry. This assumption can be formulated using the following requirements:
\begin{equation}
\pounds_{\xi^l_{(\rm a)}} U^k =0 \,, \quad a = 1,2,3,
\label{K109}
\end{equation}
i.e., the Lie derivatives of the vector field along three Killing vectors forming the Abelian subgroup vanish.
These relations require that the vector field has to depend on the retarded time only, $U^i(u)$. Also, we find automatically that (\ref{K109}) leads to
\begin{equation}
\pounds_{\xi^l_{(\rm a)}} [\nabla_i U_k] =0 \,, \quad a = 1,2,3.
\label{K119}
\end{equation}
For the metric (\ref{PW1}) the Ricci tensor has only one component $R_{uu}$, and the Ricci scalar vanishes, $R{=}0$. This means that
\begin{equation}
\xi^i_{(\alpha)} \left[R_{ik} - \frac12 R g_{ik}\right] = 0  \,,
\label{cons}
\end{equation}
leaving us with the following consequences:
$$
 \xi^i_{({\rm a})}\left[\lambda U_i U_k  {+} T^{({\rm U})}_{ik} \right] =0 \,,
$$
$$
\xi^i_{({\rm a})} \xi^k_{({\rm b})}\left[ \lambda U_i U_k  {+} T^{({\rm U})}_{ik} \right] =0 \,,
$$
 \begin{equation}
 \xi^i_{({\rm a})}U^k \left[\lambda U_i U_k  {+} T^{({\rm U})}_{ik} \right] =0 \,.
\label{cons2}
\end{equation}

\subsubsection{Aether vector field and the associated geodesic lines}

We assume that there exists a global reference frame based on the family of geodesic lines associated with the aether vector field. This means that
\begin{equation}
 \frac{d x^i}{d\tau} = U^i \,, \quad \frac{d^2 x^i}{d\tau^2} + \Gamma^i_{kl}\frac{d x^k}{d\tau}\frac{d x^l}{d\tau} = 0 \,.
\label{geodes1}
\end{equation}
Clearly, this is possible when the vector field satisfies the condition
\begin{equation}
 U^m \nabla_m U^i = 0 \,,
\label{geodes2}
\end{equation}
i.e., the acceleration vector vanishes, $DU^i{=}0$. For the metric (\ref{PW1}) the solution to Eq. (\ref{geodes2}) is known (see, e.g., \cite{Parametric}).
Indeed, (\ref{geodes2}) can be rewritten as
\begin{equation}
 U_v \partial_u U_i = \frac12 \delta_i^u g_{mn}^{\prime}(u)  U^m U^n \,,
\label{geodes21}
\end{equation}
providing the solution to be of the form
$$
 U_v = \xi^k_{(1)}U_k = {\cal E}_v \,, \quad U_2 = \xi^k_{(2)}U_k = {\cal E}_2 \,,
 $$
 \begin{equation}
 U_3 = \xi^k_{(3)}U_k = {\cal E}_3 \,, \quad U_u = \frac{1-g^{\alpha \beta} {\cal E}_{\alpha} {\cal E}_{\beta}}{2{\cal E}_v} \,,
\label{geodes3}
\end{equation}
with integration constants ${\cal E}_v$, ${\cal E}_2$ and ${\cal E}_3$.
Here and below the prime denotes the derivative with respect to
the retarded time $u$, and the Greek indices take two values $\alpha, \beta {=} 2,3$.
For such velocity four-vector, the covariant derivative is the symmetric tensor. Indeed,
$$
\Theta_{ik} = \nabla_i U_k = - \delta_i^u \delta_k^u \frac{{\cal E}_{\alpha} {\cal E}_{\beta}}{2{\cal E}_v} \left(g^{\alpha \beta}(u) \right)^{\prime}+ \frac12 {\cal E}_v g^{\prime}_{ik}(u)
-
$$
\begin{equation}
-\frac{g_{22}^{\prime}(u)}{2g_{22}} {\cal E}_2 \left(\delta_i^2 \delta_k^u {+} \delta_i^u \delta_k^2 \right)
- \frac{g_{33}^{\prime}(u)}{2g_{33}} {\cal E}_3 \left(\delta_i^3 \delta_k^u {+} \delta_i^u \delta_k^3 \right) \,,
\label{geodes4}
\end{equation}
thus the skew-symmetric vorticity tensor $\omega_{ik}$ is equal to zero identically. The corresponding expansion scalar is
\begin{equation}
\Theta \equiv g^{ik}\Theta_{ik} = 2{\cal E}_v \frac{L^{\prime}(u)}{L} \,.
\label{geodes5}
\end{equation}
Clearly, the symmetric shear tensor $\sigma_{ik}$ is also non-vanishing. For our purposes, it is sufficient to choose the constants of integration in the following form:
\begin{equation}
{\cal E}_v = \frac{1}{\sqrt2} \,, \quad {\cal E}_2 = 0 \,, \quad {\cal E}_3 = 0 \,,
\label{geodes6}
\end{equation}
providing that
\begin{equation}
U_u = U_v = \frac{1}{\sqrt2} \,, \quad U^i = \delta^i_0 \,.
\label{geodes69}
\end{equation}
This assumption means that in the chosen frame of reference the aether is in the state of rest. Now we find that the tensor $\Theta_{ik}$ has only two non-vanishing components $\Theta_{22}$ and $\Theta_{33}$:
\begin{equation}
\Theta_i^{\ k} = \frac{1}{\sqrt2} \left[\delta_i^2
\delta^k_2 \left(\frac{L^{\prime}}{L}{+}
\beta^{\prime} \right) {+} \delta_i^3 \delta^k_3
\left(\frac{L^{\prime}}{L}{-} \beta^{\prime} \right)\right]\,,
\label{GW001}
\end{equation}
and thus the quadratic invariant $\Theta_{mn} \Theta^{mn}$ takes the form
\begin{equation}
\Theta_{mn} \Theta^{mn} = \left(\frac{L^{\prime}}{L}\right)^2 + \left(\beta^{\prime}\right)^2 \,.
\label{GeoScalar}
\end{equation}
The expansion scalar reads now
\begin{equation}
\Theta = \frac{\sqrt{2}\,L^{\prime}(u)}{L} \,,
\label{GW0011}
\end{equation}
and the shear tensor can be written as
\begin{equation}
\sigma^k_i  = \frac{\Theta}{2}  \left(\frac13
\Delta_i^k - \delta_i^1 \delta^k_1 \right)
+ \frac{\beta^{\prime}}{\sqrt2} \left(\delta_i^2 \delta^k_2
{-} \delta_i^3 \delta^k_3 \right)\,.
\label{GW002}
\end{equation}

\subsubsection{Reduced equations for the vector field}

The requirement of the pp-wave symmetry in the absence of acceleration, $DU^i{=}0$, yields
\begin{equation}
I^j =  C_4 (DU_m)(\nabla^j U^m) =0 \,,
\label{RJ2}
\end{equation}
\begin{equation}
{\cal J}^{am} = \left(C_1+C_3\right) \Theta^{am} {+} C_2 g^{am} \Theta
\,. \label{2999}
\end{equation}
Then the equation $\nabla_a {\cal J}^{am} {=} \lambda U^m$ gives
\begin{equation}
\delta^m_v \left[C_2 \Theta^{\prime} {-} \sqrt2 (C_1{+}C_3) \Theta_{ab} \Theta^{ab}\right] = \frac{1}{\sqrt2}\lambda \left(\delta^m_v{+}\delta^m_u \right)\,.
\label{0A1}
\end{equation}
The two nontrivial equations (\ref{0A1}) for $m=v$ and $m=u$ are compatible if and only if $\lambda = 0$ and
\begin{equation}
(C_1{+}C_2{+}C_3) \left[\left(\frac{L^{\prime}}{L}\right)^2 {+} \left(\beta^{\prime}\right)^2 \right] = C_2 \left[ \frac{L^{\prime \prime}}{L} {+} \left(\beta^{\prime}\right)^2 \right] \,.  \label{0A309}
\end{equation}

\subsubsection{Reduced equations for the gravitational field}

For the space-time with pp-wave symmetry the Ricci scalar vanishes; in case when $\lambda=0$ we have to state that the trace of the stress-energy tensor is vanishing, $T^{({\rm U})}{=}0$.
Thus, we obtain from (\ref{trace})
$$
2 (C_1+C_3) \Theta^{am}\Theta_{am} {-}
\frac{1}{\sqrt2}\left(C_1+4C_2+C_3 \right)\Theta^{\prime} -
$$
\begin{equation}
-
\left(C_1+2C_2+C_3 \right)\Theta^2 =0 \,,
\label{trace2}
\end{equation}
or in more detail
$$
(C_1+C_2+C_3) \left[2\left(\frac{L^{\prime}}{L}\right)^2 + 3\left(\beta^{\prime}\right)^2 \right] +C_2 \left(\beta^{\prime}\right)^2 =
$$
\begin{equation}
=
\left(C_1+4C_2+C_3 \right)\left[\frac{L^{\prime \prime}}{L} + \left(\beta^{\prime}\right)^2 \right]\,.
\label{trace3}
\end{equation}
Combination of the two equalities (\ref{trace3}) and (\ref{0A309}) requires that
\begin{equation}
C_1+C_3 =0 \,, \quad C_2 =0  \ \Rightarrow  {\cal C}_{\sigma} = 0 \,, \quad {\cal C}_{\Theta}= 0 \,.
\label{trace44}
\end{equation}
In other words, we can consider the aether to possess pure pp-wave modes, if and only if the velocity of tensorial mode coincides with the speed of light in vacuum, $S_{(2)}{=}1$, and the scalar modes are stopped, $S_{(0)}=0$ (see (\ref{act951}) and (\ref{act952})).
In this case we immediately obtain from (\ref{299}) and (\ref{5Ein1})
\begin{equation}
{\cal J}^{am} =0 \,, \quad T_{ik}^{({\rm U})}=0 \,.
\label{trace4}
\end{equation}
Thus, the gravity field equations reduce to one equation, $R_{uu}{=}0$, which has the well-known form
\begin{equation}
\frac{L^{\prime \prime}}{L} + {\beta^{\prime}}^2 = 0 \,. \label{act53}
\end{equation}
The parameters $C_4$ and $C_1{-}C_3{=}C_{\omega}$ remain hidden parameters of the model.

\subsubsection{Resume: definition of the aether pp-wave mode}

To conclude, we can define the {\it aether pp-wave mode} as a state of the $\left\{g,U \right\}$ field configuration, for which the metric $g_{ik}$ (\ref{PW1}) relates to $G_5$ group of isometries, satisfies Eq. (\ref{act53}), and for which the unit vector field is characterized by Eqs. (\ref{geodes69}) and (\ref{GW001}). This state of the aether, the pp-wave mode, is presented by the {\it exact solution} to the total coupled system of equations for the vector and gravity fields in the framework of the truncated Einstein-aether model with two arbitrary coupling constants ($C_1$ and $C_4$)
and two fixed ones ($C_2{=}0$ and $C_3 {=} {-}C_1$). In the approximation of weak fields, this exact solution corresponds to the particular case of the Einstein-aether waves \cite{J2}, for which the scalar mode is stopped and the tensorial mode propagates with the velocity equal to the speed of light in the standard vacuum.

\subsection{Extended theory including the Maxwell field}

\subsubsection{Extended action functional}

In \cite{BL2014} the Einstein-aether theory was extended by including all admissible terms with the Maxwell tensor $F_{ik}$. Now we consider a particular Einstein--Maxwell-aether model, which is based on the action functional
\begin{equation}
{\cal S}_{({\rm total})} = {\cal S}_{(0)}+ {\cal S}_{({\rm EMA})}\,,
\label{EMAaction1}
\end{equation}
where the additional functional is of the form
$$
{\cal S}_{({\rm EMA})} = \frac{1}{4} \int d^4 x \sqrt{{-}g} \left[F^{mn}F_{mn} +  \right.
$$
\begin{equation}
\left.
+X^{pqikmn} \nabla_p U_q F_{ik}F_{mn} \right] \,.
\label{M20}
\end{equation}
The tensor $X^{pqikmn}$ describes the coupling of electromagnetic field to the non-uniformly moving aether; it was reconstructed in \cite{BL2014} using the metric $g_{ik}$, the covariant constant Kronecker tensors
($\delta^i_k$, $\delta^{ik}_{ab}$ and higher order Kronecker tensors),
the Levi-Civita tensor $\epsilon^{ikab}$, and the unit vector field $U^k$ itself. In the context of our study we extract the two-parameter version of the tensor $X^{pqikmn}$:
$$
X^{lsikmn} =
 \frac14 U_p U_q \left[\alpha  \left(g^{iklp}g^{mnsq} +
g^{mnlp}g^{iksq} \right)-  \right.
$$
\begin{equation}
\left.
- \gamma \
\left(\epsilon^{iklp} \epsilon^{smnq} + \epsilon^{iksp} \epsilon^{lmnq} \right)\right]
\,.
\label{X}
\end{equation}
This tensor contains two coupling constants, $\alpha$ and $\gamma$, describing the dynamo-optical interactions. Comparing this version with the total one presented in \cite{BL2014}, one can see that we put
\begin{equation}
\alpha = \alpha_6 = 3 \alpha_1 \,, \quad \gamma = \gamma_6 = 3 \gamma_1 \,.
\label{X99}
\end{equation}
Also, we use the auxiliary tensor
\begin{equation}
g^{ikmn} \equiv g^{im}g^{kn}{-}g^{in}g^{km} \,.
\label{X79}
\end{equation}
As follows from \cite{BL2014}, this choice of the set of phenomenological parameters relates to the following extensions of the permittivity tensors:
\begin{equation}
\varepsilon^{ik} {=} \Delta^{ik} {+} \alpha \Theta^{ik}  \,,  {\left(\mu^{{-}1}\right)}^{ik} {=} \Delta^{ik} {+}
\gamma \Theta^{ik}  \,, \quad \nu^{ik} {=} 0 \,. \label{eps}
\end{equation}
Here  $\Delta^{ik}=g^{ik}-U^iU^k$ is the projector,  and $\alpha$ and $\gamma$ are two new independent dynamo-optical coupling constants. The term $\alpha \Theta^{ik}$ describes electric susceptibility induced in the aether by the pp-wave mode; the term $\gamma \Theta^{ik}$ relates to the inverse magnetic susceptibility. When the aether flow is uniform, $\Theta^{ik}=0$, the coupling of photons to the aether remains latent, thus the parameters $\alpha$ and $\gamma$ are hidden.
It seems to be interesting to mention that generally the tensors $\varepsilon^{ik}$ and ${\left(\mu^{-1}\right)}^{ik}$ become anisotropic, when $\Theta^{ik} \neq 0$. Taking into account (\ref{GW001}) one can say that these tensors have three different eigen-values and the electromagnetically active aether behaves as a bi-axial quasi-medium.

\subsubsection{Electrodynamic equations}

The extended system of electrodynamic equations contains two subsets \cite{BL2014}:
\begin{equation}
\nabla_k \left[F^{ik}+ X^{pqikmn} \nabla_p U_q F_{mn} \right] = 0  \,,
\label{E5}
\end{equation}
\begin{equation}
\nabla_k F^{*ik} = 0 \,.
\label{E51}
\end{equation}
The first subset is the result of variation of the extended action functional (\ref{EMAaction1}) with respect to the potential of the electromagnetic field $A_i$, which defines the Maxwell tensor
\begin{equation}
F_{ik} = \nabla_i A_k - \nabla_k A_i\,.
\label{maxwellA1}
\end{equation}
The second subset is the standard consequence of (\ref{maxwellA1}) written in terms of the dual Maxwell tensor
$F^{*ik} \equiv \frac12 \epsilon^{ikmn}F_{mn}$ ($\epsilon^{ikmn}=\frac{1}{\sqrt{-g}} E^{ikmn}$ with $E^{0123}=1$).
As usual, we consider the Lorentz gauge for the potential four-vector, $\nabla_k A^k =0$. Now we are ready for extended analysis of the solutions to the electrodynamic equations.

\section{Analysis of solutions to the electrodynamic equations in the model with the pp-wave symmetry of the aether flow}

\subsection{Preamble}

\subsubsection{General solutions for the electromagnetic waves in the absence of coupling to the aether pp-wave modes}

Let us recall, first, the results known for the model with the pp-wave symmetry for the case when the photon--aether interactions are absent, i.e., $\alpha{=}0$, $\gamma{=}0$.
For this purpose we can use the results of Ref. \cite{B97}, which describes the influence of a gravitational pp-wave on the electromagnetic wave with arbitrary direction of propagation. Of course, in that work we did not consider the aether, however, the basic gravity field equation was used in the form coinciding with (\ref{act53}). As shown in \cite{B97} for the minimal case, the equations for the components $A_u$, $A_v {=} \xi^i_{(1)}A_i$, $A_2 {=} \xi^i_{(2)}A_i$, $A_3 {=} \xi^i_{(3)}A_i$ of the electromagnetic potential four-vector can be decoupled taking three steps.

\noindent
{\it First}, the equation for the component $A_v$ is self-closed,
\begin{equation}
{\cal D} A_v +2\frac{L^{\prime}}{L} \partial_v A_v = 0\,,
\label{old1}
\end{equation}
where we use the following operator
\begin{equation}
{\cal D} \equiv 2 \partial_u \partial_v + g^{\alpha \beta}(u) \partial_{\alpha} \partial_{\beta} \,.
\label{old2}
\end{equation}
The solution to this equation is
\begin{equation}
A_v = \frac{1}{L} B_v(W) + B_v^{*}(u) \,,
\label{old30}
\end{equation}
\begin{equation}
 W = W_0 + k_v v  + k_2 x^2 +k_3 x^3 - \frac{k_{\alpha} k_{\beta}}{2k_v} \int_0^u dz g^{\alpha \beta}(z)\,.
\label{old3}
\end{equation}
Here $B_v(W)$ is an arbitrary function of its argument, the phase function $W$; the constants $k_v$, $k_2$, $k_3$ play the roles of components of the corresponding wave vector; $W_0$ is the constant of integration; $B_v^{*}(u)$ is an arbitrary function of the retarded time $u$.
The wave four-vector
\begin{equation}
K_m \equiv \nabla_m W = {-} \delta_m^u \frac{k_{\alpha} k_{\beta}}{2k_v} g^{\alpha \beta}(u) {+} \delta_m^v k_v   {+} \delta_m^2 k_2  {+} \delta_m^3 k_3 \,,
\label{old5}
\end{equation}
is the null four-vector, $K_mK^m {=0}$.

\noindent
{\it Second}, the equations for the components $A_2$ and $A_3$ contain the component $A_v$ already known:
\begin{equation}
{\cal D} A_2 - 2 \beta^{\prime}\partial_v A_2  + 2\left(\frac{L^{\prime}}{L}+ \beta^{\prime}\right) \partial_2 A_v =0 \,,
\label{old6}
\end{equation}
\begin{equation}
{\cal D} A_3 + 2 \beta^{\prime}\partial_v A_3 + 2\left(\frac{L^{\prime}}{L}- \beta^{\prime}\right) \partial_3 A_v=0 \,.
\label{old7}
\end{equation}
The corresponding solutions are
\begin{equation}
A_2 = e^{\beta(u)} B_2(W) +\frac{k_2}{k_v L} B_v(W) + B^{*}_2(u)  \,,
\label{old8}
\end{equation}
\begin{equation}
A_3 = e^{-\beta(u)}B_3(W) + \frac{k_3}{k_v L} B_v(W)  + B^{*}_3(u) \,,
\label{old9}
\end{equation}
where $B_2(W)$ and $B_3(W)$ are arbitrary functions of the phase $W$ given by (\ref{old3}), and $B^{*}_2(u)$, $B^{*}_3(u)$ are arbitrary functions of the retarded time only.

\noindent
{\it Third}, we obtain the component $A_u$ from the Lorentz gauge condition $\nabla_kA^k {=}0$ rewritten as
\begin{equation}
\partial_v A_u +\partial_u A_v  - \frac{e^{-2\beta}}{L^2} \partial_2 A_{2} - \frac{e^{2\beta}}{L^2} \partial_3 A_{3} + 2A_v \frac{L^{\prime}}{L} =0\,.
\label{old11}
\end{equation}
The corresponding solution is of the form
$$
A_u = B_u^{*}(u) + \frac{1}{k_v L^2}[e^{-\beta}k_2 B_2(W) {+} e^{\beta}k_3 B_3(W)]{-}
$$
$$
-\frac{v}{L^2}\frac{d}{du}\left[L^2 B_v^{*}(u)\right] {-} \frac{L^{\prime}}{k_v L^2} \int dW B_v(W)  +
$$
\begin{equation}
 {+}
\frac{1}{2k^2_v L^3}B_v(W)\left[k^2_2 e^{-2\beta}{+} k^2_3 e^{2\beta}\right]\,.
\label{old12}
\end{equation}
Below we extract two examples from the general solution presented in this section, since they will play the important role in our further analysis.

\subsubsection{Basic example I: longitudinal configuration}

Let the guiding parameters and arbitrary functions be chosen as follows: $k_2{=}k_3{=}0$, $W_0 {=}0$, and $B_v(W)=0$, $B^{*}_v(u) = B^{*}_u(u)=0$. Then we find from (\ref{old30}), (\ref{old3}), (\ref{old8}), (\ref{old9}), (\ref{old12}) that
$$
A_2 = e^{\beta(u)} B_2(k_v v) + B^{*}_2(u)  \,,
$$
\begin{equation}
A_3 = e^{-\beta(u)}B_3(k_v v) + B^{*}_3(u) \,, \quad A_v = A_u =0 \,.
\label{old779}
\end{equation}
This solution describes the electromagnetic waves with the front parallel to the gravitational wave front. The so-called co-moving electromagnetic wave is described by two arbitrary functions of the retarded time,
\begin{equation}
A_2 = B^{*}_2(u)  \,, \quad A_3 = B^{*}_3(u) \,.
\label{old579}
\end{equation}
If the electromagnetic wave propagates towards the gravitational wave, we deal, respectively, with the potentials
\begin{equation}
A_2 = e^{\beta(u)} B_2(k_v v) \,, \quad A_3 = e^{-\beta(u)}B_3(k_v v)  \,.
\label{old179}
\end{equation}
Clearly, for this wave configuration there is no birefringence induced by a pp-wave gravitational field \cite{B97}.

\subsubsection{Basic example II: transversal configuration}

Let arbitrary functions be chosen so that
$$
B_2(W){=}0 \,, \quad B^{*}_v(u) {=}0 \,,  \quad  B^{*}_2(u){=}0 \,,
$$
\begin{equation}
B^{*}_3(u){=}0\,, \quad B_u^{*}(u){=}0 \,.
\label{old55}
\end{equation}
In addition, we choose the constants $k_v, k_2, k_3$ as follows:
\begin{equation}
k_v = \frac{k}{\sqrt2} \,, \quad k_2 = -k \,, \quad k_3 = 0 \,.
\label{old59}
\end{equation}
Then we find immediately from (\ref{old30}), (\ref{old3}), (\ref{old8}), (\ref{old9}), and (\ref{old12}) that
\begin{equation}
 A_u =  \frac{1}{L^3}B_v(W)  e^{-2\beta} - \frac{\sqrt2 L^{\prime}}{kL^2} \int dW B_v(W)\,,
\label{old54}
\end{equation}
\begin{equation}
 A_v = \frac{1}{L} B_v(W) \,, A_2 = {-} \frac{\sqrt2}{L} B_v(W) \,, A_3 = e^{-\beta}B_3(W) \,,
\label{old541}
\end{equation}
\begin{equation}
W = W_0 {+} k \left[\frac{v{+}u}{\sqrt2} {-} x^2 {+} \frac{1}{\sqrt2 }\int_0^u du \left(\frac{e^{-2\beta}}{L^{2}} {-}1\right) \right].
\label{old524}
\end{equation}
When $\beta=0$ and $L=1$, the phase (\ref{old524}) transforms into $W = W_0 + k (t - x^2)$, and we deal with the electromagnetic wave propagating along $0x^2$. The first polarization of this electromagnetic wave is associated with the direction $0x^3$ and is described by the function $B_3(W)$; the second polarization is described by the function $B_v(W)$. Thus, in the field of pure gravitational wave (without aether), there is no birefringence, since the phase $W$ relates to both polarizations.

\subsection{Analysis of solutions to the extended electrodynamic equations}

\subsubsection{Equations to be solved}

When the phenomenological coupling parameters $\alpha$ and $\gamma$ are non-vanishing, we have to solve the extended system of electrodynamic equations. To be more precise, instead of Eqs.
(\ref{old1}), (\ref{old6}), and (\ref{old7}) we deal, respectively, with the equations
\begin{equation}
{\cal D} A_v + 2\frac{L^{\prime}}{L} \partial_v A_v = X_v \,,
\label{new1}
\end{equation}
\begin{equation}
{\cal D} A_2 - 2 \beta^{\prime}\partial_v A_2  + 2\left(\frac{L^{\prime}}{L}+ \beta^{\prime}\right) \partial_2 A_v = X_2 \,,
\label{new6}
\end{equation}
\begin{equation}
{\cal D} A_3 + 2 \beta^{\prime}\partial_v A_3 + 2\left(\frac{L^{\prime}}{L}- \beta^{\prime}\right) \partial_3 A_v =X_3 \,,
\label{new7}
\end{equation}
and only Eq. (\ref{old11}) remains unchanged. The new terms in the right-hand sizes of Eqs. (\ref{new1}), (\ref{new6}), and  (\ref{new7}) have, respectively, the form
\begin{equation}
X_v = X^{\alpha \beta u\sigma mn} \Theta_{\alpha \beta} \partial_{\sigma}F_{mn} \,,
\label{new15}
\end{equation}
$$
X_2 \equiv -e^{2\beta}\left[\partial_u \left(L^2 X^{ls2umn}\Theta_{ls}F_{mn} \right)+ \right.
$$
\begin{equation}
\left.
+\partial_v \left(L^2 X^{ls2vmn}\Theta_{ls}F_{mn} \right) \right] \,,
\label{new65}
\end{equation}
$$
X_3 \equiv -e^{-2\beta} \left[\partial_u \left(L^2 X^{ls3umn}\Theta_{ls}F_{mn} \right)+ \right.
$$
\begin{equation}
\left.
+\partial_v \left(L^2 X^{ls3vmn}\Theta_{ls}F_{mn} \right) \right] \,.
\label{new75}
\end{equation}
In more detail, we can present these terms as follows:
$$
X_v = \frac{e^{-2\beta}}{L^2} \partial^2_2  \left(h_1 A_v {+} h_2 A_u \right)
{+} \frac{e^{2\beta}}{L^2} \partial^2_3  \left(h_3 A_v {+} h_4 A_u \right) {-}
$$
\begin{equation}
{-} \frac{e^{-2\beta}}{L^2} \left(h_1 \partial_v {+} h_2 \partial_u \right) \partial_2 A_2
- \frac{e^{2\beta}}{L^2} \left(h_3 \partial_v {+} h_4 \partial_u \right) \partial_3 A_3
 \,,
\label{new16}
\end{equation}
$$
X_2 = - 2 h_1 \partial_u \partial_v A_2 - h_2 \left(\partial^2_u + \partial^2_v \right)A_2 +
$$
$$
+ 2\beta^{\prime} \left(h_1 \partial_v + h_2 \partial_u \right) A_2 -
\left(h^{\prime}_1 \partial_v + h^{\prime}_2 \partial_u \right) A_2 +
$$
$$
\partial_2 \left\{h_1\left[\partial_u A_v + \partial_v A_u \right] + h_2\left[\partial_u A_u + \partial_v A_v \right]-
\right.
$$
\begin{equation}
\left.
- 2\beta^{\prime}\left[h_1A_v + h_2 A_u \right]
+ \left[h_1^{\prime}A_v + h_2^{\prime}A_u \right]\right\}
\,,
\label{new162}
\end{equation}
$$
X_3 = - 2 h_3 \partial_u \partial_v A_3 - h_4 \left(\partial^2_u + \partial^2_v \right)A_3 -
$$
$$
- 2\beta^{\prime} \left(h_3 \partial_v + h_4 \partial_u \right) A_3 -
\left(h^{\prime}_3 \partial_v + h^{\prime}_4 \partial_u \right) A_3 +
$$
$$
\partial_3 \left\{h_3\left[\partial_u A_v + \partial_v A_u \right] + h_4\left[\partial_u A_u + \partial_v A_v \right]+ \right.
$$
\begin{equation}
\left.
+2\beta^{\prime}\left[h_3 A_v + h_4 A_u \right]
+ \left[h_3^{\prime}A_v + h_4^{\prime}A_u \right]\right\}
\,.
\label{new17}
\end{equation}
The auxiliary functions $h_1(u)$, $h_2(u)$, $h_3(u)$, $h_4(u)$ are
$$
h_1(u) = \frac{1}{2\sqrt2}\left[\frac{L^{\prime}}{L}(\alpha+\gamma)+ \beta^{\prime}(\alpha-\gamma) \right] \,,
$$
$$
h_2(u) = \frac{1}{2\sqrt2}\left[\frac{L^{\prime}}{L}(\alpha-\gamma)+ \beta^{\prime}(\alpha+\gamma) \right] \,,
$$
$$
h_3(u) = \frac{1}{2\sqrt2}\left[\frac{L^{\prime}}{L}(\alpha+\gamma)- \beta^{\prime}(\alpha-\gamma) \right] \,,
$$
\begin{equation}
h_4(u) = \frac{1}{2\sqrt2}\left[\frac{L^{\prime}}{L}(\alpha-\gamma)- \beta^{\prime}(\alpha+\gamma) \right] \,.
\label{newh}
\end{equation}
Since the term $X_v$ contains $A_u$, $A_2$, $A_3$, when the interaction of electromagnetic waves with the pp-wave aether modes exists, the component $A_v$ is not decoupled, as in the case $\alpha {=} \gamma {=}0$ (see (\ref{old1})). Now the solution is much more sophisticated.

\subsubsection{Electromagnetic waves with the front parallel to the front of the pp-wave aether mode}

We start the analysis with the case for which the solutions depend neither on $x^2$, nor on $x^3$. The corresponding electromagnetic wave propagates along $x^1$.
For such model we obtain from (\ref{new16}) that $X_v{=}0$. Consequently, $A_v{=}0$ is the solution to Eq. (\ref{new1}), and  $A_u{=}0$ is now the solution to (\ref{old11}).
As a result, we are faced with two decoupled equations for two unknown functions $A_2$ and $A_3$:
$$
2\left(1+h_1 \right)\partial_u \partial_v A_2 + h_{2} \left[\partial^2_u + \partial^2_v \right]A_2 =
$$
\begin{equation}
=\left(2\beta^{\prime} h_2 - h_2^{\prime} \right) \partial_u A_2 + \left[2\beta^{\prime} (1+h_1)-h_1^{\prime} \right]\partial_v A_2 \,,
\label{nch69}
\end{equation}
$$
2\left(1+h_3 \right)\partial_u \partial_v A_3 + h_{4} \left[\partial^2_u + \partial^2_v \right]A_3 = $$
\begin{equation}
-\left(2\beta^{\prime} h_4 + h_4^{\prime} \right) \partial_u A_3 - \left[2\beta^{\prime} (1+h_3)+h_3^{\prime} \right]\partial_v A_3 \,.\label{nch79}
\end{equation}
Let us focus on the first one, and make the following remark concerning the co-moving electromagnetic wave: when $h_2(u)=0$, there exists the solution $A_2(u)=B^{*}_2(u)$, where $B^{*}_2(u)$ is an arbitrary function of the retarded time; when $h_2(u) \neq 0$, an arbitrary co-moving wave is not admissible, there is only very specific solution of this type, namely $A_2(u)= K_1+K_2\int du e^{2\beta} h^{-1}_2$. From the physical point of view, this means that the phase velocity of the electromagnetic wave in the aether differs from the speed of light in vacuum, and thus, both retarded and advanced times, $u$ and $v$, have to form the argument of the potential, $A_2(u,v)$.
The equation of the characteristics associated with (\ref{nch69}) is of the form
\begin{equation}
h_{2} \left(dv^2 + du^2 \right) -2\left(1+h_1 \right)du dv  = 0 \,.
\label{ch1}
\end{equation}
When $h_2{=}0$, this equation converts into $du dv {=}0$, providing the characteristics to be in the form $u=const$, $v=const$. When $h_2(u) \neq 0$, two first integrals of Eq. (\ref{ch1}) are
\begin{equation}
v - \int \frac{du}{H_{21}}\left(1 \pm \sqrt{1-H^2_{21}}\right) = K_{(\pm)} \,,
\label{2ch3}
\end{equation}
where the auxiliary function is introduced as
\begin{equation}
H_{21}(u) = \frac{h_2(u)}{1+h_1(u)} \,.
\label{ch3}
\end{equation}
Equation (\ref{nch69}) is of the hyperbolic type, when $|H_{21}|<1$.
Thus we obtain two characteristics in the form
$$
\xi_2 \equiv v - \int \frac{du}{H_{21}}\left(1+ \sqrt{1-H^2_{21}}\right)  \,,
$$
\begin{equation}
\eta_2 \equiv v - \int \frac{du}{H_{21}}\left(1- \sqrt{1-H^2_{21}}\right) \,.
\label{ch5}
\end{equation}
(The superscript $2$ in $\xi_2$ and $\eta_2$ relates to the polarization along $0x^2$). In the limit $\alpha \to 0$, $\gamma \to 0$, we obtain that $\eta_2 \to v$. As for the first characteristic, in the limit $\alpha \to 0$, $\gamma \to 0$ the integral tends to infinity; this means that there is no continuous transition between characteristics $u=const$ and $\xi_2=const$.

The same result can be obtained using the eikonal equation. Indeed, when $A_2 \to a_2 e^{i[k_v v+ \sigma(u)]}$ with large phase and slowly varying amplitude $a_2$, we obtain the following leading order equation:
\begin{equation}
2\left(1+h_1 \right) k_v \sigma^{\prime}  + h_{2} \left({\sigma^{\prime}}^2 + k_v^2 \right) = 0 \,.
\label{ch69}
\end{equation}
Clearly, the solution to this equation,
\begin{equation}
{\sigma}^{\prime}_{(\pm)} = \frac{k_v}{H_{21}}\left(-1 \pm \sqrt{1-H^2_{21}}\right)
\label{4ch3}
\end{equation}
covers the results displayed in (\ref{ch5}).
When we consider Eq. (\ref{nch79}) searching for the component $A_3$, we obtain similar results:
$$
\xi_3 \equiv v - \int \frac{du}{H_{31}}\left(1+ \sqrt{1-H^2_{31}}\right)  \,,
 $$
 \begin{equation}
 \eta_3 \equiv v - \int \frac{du}{H_{31}}\left(1- \sqrt{1-H^2_{31}}\right) \,,
\label{ch56}
\end{equation}
where the function
\begin{equation}
H_{31}(u) = \frac{h_4(u)}{1+h_3(u)} \,,
\label{h39}
\end{equation}
can be obtained from $H_{21}(u)$ with the replacement $\beta \to - \beta$.
Clearly, the phases of two components, $A_2$ and $A_3$ do not coincide, we deal with the manifestation of birefringence.

\subsubsection{Searching for an analog of transversal electromagnetic wave}

Our assumption is now that the electromagnetic potentials do not depend on the variable $x^3$. We are faced now with the fact that the system of master equations splits into two subsystems.
The equation for the component $A_3$ is decoupled and has the following form:
$$
2\left(1+h_3 \right)\partial_u \partial_v A_3 + h_{4} \left[\partial^2_u + \partial^2_v \right]A_3 - \frac{e^{-2\beta}}{L^2} \partial^2_2 A_3=
 $$
 \begin{equation}
 = -\left(2\beta^{\prime} h_4 + h_4^{\prime} \right) \partial_u A_3 - \left[2\beta^{\prime} (1+h_3)+h_3^{\prime} \right]\partial_v A_3 \,.\label{5nch79}
\end{equation}
The equations for $A_u$, $A_v$ and $A_2$ remain coupled
\begin{equation}
\partial_v A_u +\partial_u A_v - \frac{e^{-2\beta}}{L^2}\partial_2 A_2 + 2A_v \frac{L^{\prime}}{L} =0\,,
\label{t1}
\end{equation}
$$
2 \partial_u \partial_v A_v - \frac{e^{-2\beta}}{L^2} \left( 1+h_1 \right) \partial^2_2 A_v + 2\frac{L^{\prime}}{L} \partial_v A_v =
$$
\begin{equation}
= \frac{e^{-2\beta}}{L^2}  \left[h_2 \partial^2_{2}A_u - (h_1 \partial_v + h_2 \partial _u)\partial_2 A_2 \right]\,,
\label{t2}
\end{equation}
$$
2(1+h_1)\partial_u \partial_v A_2 - \frac{e^{-2\beta}}{L^2} (1+h_1)\partial^2_2 A_2 +
$$
$$
+ h_{2} \left[\partial^2_u + \partial^2_v \right]A_2 - h_2 \partial_2\left(\partial_u A_u + \partial_v A_v \right)=
$$
$$
= (2\beta^{\prime}h_2 -h_2^{\prime}) \partial_u A_2 + \left[2\beta^{\prime}(1+h_1)-h_1^{\prime} \right]\partial_v A_2-
$$
\begin{equation}
{-} \left[2(1{+}h_1)\left(\frac{L^{\prime}}{L}{+} \beta^{\prime} \right){-} h_1^{\prime} \right]\partial_2 A_v {-} (2\beta^{\prime}h_2 {-} h_2^{\prime}) \partial_2 A_u
\,.
\label{t3}
\end{equation}
In order to study the birefringence effect, we can use the eikonal approximation and consider  the potentials as follows (compare with (\ref{old54})--(\ref{old524})):
$$
A_u \to a_u e^{i\Psi} \,,  A_v \to a_v e^{i\Psi} \,, A_2 \to a_2 e^{i\Psi} \,, A_3 \to a_3 e^{i\Phi} \,,
$$
\begin{equation}
\Psi = k \left[\frac{v}{\sqrt2} {-} x^2 {+} \psi(u) \right] \,, \quad \Phi = k \left[\frac{v}{\sqrt2} {-} x^2 {+} \phi(u) \right] \,.
\label{t4}
\end{equation}
We analyze, first, the decoupled equation (\ref{5nch79}) for the component $A_3$. The eikonal equation
\begin{equation}
h_4\left({\phi^{\prime}}^2+ \frac12 \right) + \sqrt2 (1+h_3)\phi^{\prime} - \frac{e^{-2\beta}}{L^2}=0
\label{t5}
\end{equation}
gives two phase functions
\begin{equation}
\phi_{(\pm)}(u) = \int_0^u  \frac{du}{\sqrt2 H_{31}} \left[{-}1 \pm \sqrt{1{+}H_{31}^2 \left(2\frac{e^{{-}2\beta}}{h_4 L^2}{-}1 \right)} \  \right].
\label{t7}
\end{equation}
When we apply the eikonal formalism to the coupled system of equations for $A_u$, $A_v$, $A_2$, (\ref{t1}), (\ref{t2}), (\ref{t3}), we are faced with the algebraic system, which it is convenient to present in the following "$1{+}2$" form:
$$
a_u = - \sqrt2 \left[\psi^{\prime} a_v + \frac{e^{-2\beta}}{L^2} a_2 \right] \,,
$$
$$
a_v \left[1+ h_1 - \sqrt2 L^2 e^{2\beta} \psi^{\prime} - \sqrt2 h_2 \psi^{\prime}\right] +
$$
$$
+a_2 \left[\frac{1}{\sqrt2}h_1 + \psi^{\prime} h_2 - \sqrt2 h_2 \frac{e^{-2\beta}}{L^2}\right] = 0 \,,
$$
$$
a_v \frac{h_2}{\sqrt2}\left[1-2{\psi^{\prime}}^2 \right] + a_2 \left[h_2 \left({\psi^{\prime}}^2 + \frac12 - \sqrt2 \psi^{\prime} \frac{e^{-2\beta}}{L^2} \right) + \right.
 $$
 \begin{equation}
\left.
 +(1+h_1)\left(\sqrt2 \psi^{\prime} - \frac{e^{-2\beta}}{L^2} \right)\right] =0 \,.
\label{t8}
\end{equation}
The system of last two equations admits a nontrivial solution, when the Cramer determinant vanishes, and the function $\psi^{\prime}$ satisfies the following cubic equation:
$$
-{\psi^{\prime}}^3 \left[\sqrt2 h_2L^2 e^{2\beta} \right] + {\psi^{\prime}}^2 \left[h_2 - 2(1+h_1) L^2 e^{2\beta}\right] +
$$
$$
+ \psi^{\prime} \sqrt2 \left[(1+h_1)(2+h_1) - h_2^2 - \frac12 h_2 L^2 e^{2\beta}\right] +
$$
\begin{equation}
+\left\{\frac12 h_2 + \frac{e^{-2\beta}}{L^2}\left[h_2^2 -(1+h_1)^2\right] \right\}=0 \,.
\label{t9}
\end{equation}
When the electromagnetic wave does not interact with the aether, i.e., $\alpha{=}\gamma {=}0$, this cubic equation transforms into the quadratic one
\begin{equation}
\left[\sqrt2 \psi^{\prime} L e^{\beta} -\frac{e^{-\beta}}{L}\right]^2 =0 \,,
\label{t10}
\end{equation}
thus providing the solution to be the double and to have the form (compare with (\ref{old524}))
\begin{equation}
\psi^{\prime}_{(0)} =  \frac{e^{-2\beta}}{\sqrt2 L^2} \,.
\label{t11}
\end{equation}
When $\alpha \neq 0$ and/or $\gamma \neq 0$, surprisingly, the root $\psi^{\prime}_{(0)}$ (\ref{t11}) is again one of the roots of cubic equation (\ref{t9}), and for this root
we obtain
$a_u =  a_v \frac{e^{-2\beta}}{L^2}$, and $a_2 = -\sqrt2 a_v$.

The other two roots of the cubic equation, and  the corresponding amplitudes, can be written as follows:
\begin{equation}
\psi^{\prime}_{(\pm)} = \frac{1}{\sqrt2 H_{21}}\left\{-1 \pm \sqrt{\left(1-H^2_{21} \right) \left[1+ 2h_2 \frac{e^{-2\beta}}{L^2} \right]} \right\}  \,,
\label{t12}
\end{equation}
\begin{equation}
a^{(\pm)}_2 = - \sqrt2  a_v \frac{\left[1+ h_1 - \sqrt2 \psi^{\prime}_{(\pm)} \left(L^2 e^{2\beta}  + h_2 \right)\right]}{\left[h_1 + \sqrt2  h_2 \left(\psi^{\prime}_{(\pm)}  - \sqrt2 \frac{e^{-2\beta}}{L^2}\right)\right]}       \,,
\label{t13}
\end{equation}
$$
a^{(\pm)}_u =  \sqrt2 a_v  \left[h_1 + \sqrt2 h_2\left( \psi^{\prime}_{(\pm)} - \sqrt2 \frac{e^{-2\beta}}{L^2}\right) \right]^{-1} \ \times
$$
\begin{equation}
 \left[\sqrt2 \frac{e^{-2\beta}}{L^2}(1{+}h_1) {-} \sqrt2 h_2 {\psi^{\prime}_{(\pm)}}^2 {-} (2{+}h_1)\psi^{\prime}_{(\pm)} \right]\,.
\label{t14}
\end{equation}
Based on the solutions obtained we consider below the birefringence effect and symptoms of anomalous behavior of the electromagnetic response.

\section{Discussion}

\subsection{Phase velocities of polarized electromagnetic waves in the excited aether}

We have shown that the dynamical aether, being  excited by the pp-wave modes, can manifest properties of bi-axial medium, when the test electromagnetic wave propagates in the aether.
A possible anisotropy of the dielectric and magnetic susceptibility tensors can appear in a dynamo-optical manner, i.e., it can be produced by the shear and expansion of the aether flow.
Since the pp-wave aether mode is considered as a provider of such anisotropy, we can identify three eigen-axes with the direction of the aether mode propagation ($0x^1$) as well as with directions of the first and second polarizations of the tensorial mode ($0x^2$ and $0x^3$, in our model).  For instance, when we deal with the dielectric properties of the aether with the metric (\ref{PW1}) and Killing vectors (\ref{K1}), we can use the ellipsoid
$$
\frac{\left({x^1}\right)^2}{\varepsilon_1} + \frac{\left({x^2}\right)^2}{\varepsilon_2} + \frac{\left({x^3}\right)^2}{\varepsilon_3} =1 \,,
$$
\begin{equation}
 \varepsilon_1 =1 ,  \quad \varepsilon_2 =1{+} \alpha \left(\frac{L^{\prime}}{L}{+}\beta^{\prime}\right),  \varepsilon_3 =1{+} \alpha \left(\frac{L^{\prime}}{L}{-}\beta^{\prime}\right),
\label{eps3}
\end{equation}
as the characteristic surface of this quasi-medium (see, e.g., \cite{Sirotin}). Clearly, when $\beta(u) \neq 0$, there are no coinciding eigen-values, $\varepsilon_1 \neq \varepsilon_2 \neq \varepsilon_3$. In the classical electrodynamics of bi-axial media, there is a prediction as regards the existence of {\it one} ordinary and {\it two} extraordinary waves. In fact, when the wave falls on the boundary of the bi-axial medium, it splits into two waves displaying the birefringence effect, however, the pair of waves (one ordinary plus one extraordinary) can be realized in two variants. These waves are characterized by different phase velocities, which depend on the direction of the wave propagation and on the wave polarization. In order to define the phase velocity we calculate, first, the wave four-vector $K_i$ as a four-gradient of the phase $K_i = \nabla_i W$, and its square $K_iK^i$; then we find the wave frequency as the projection of this four-vector on the aether velocity four-vector, $\Omega = K_i U^i$ (for definiteness, we choose the positive value of this projection in all cases considered below). The next step in our procedure (see \cite{BKL}) is the calculation of the spatial wave four-vector $K_i^{*}= \Delta_i^j K_j$, and of its modulus $K^{*}$. Finally, we obtain the phase velocity according to the rule
$V_{({\rm ph})} = \frac{\Omega}{\ K^{*}}$. Let us apply the described procedure to both longitudinal and transversal submodels.

\subsubsection{Phase velocities  of the electromagnetic waves with the front parallel to the pp-wave front}

Taking into account Eqs. (\ref{ch5}),  (\ref{4ch3}), and (\ref{ch56}), we see that
$$
K_i = \delta_i^{u}\sigma^{\prime} + \delta_i^{v} k_v \,, \quad K_m K^m = 2 k_v \sigma^{\prime} \,,
$$
$$
\Omega = K_i U^i = \frac{1}{\sqrt2} \left(k_v + \sigma^{\prime} \right) \,,
$$
\begin{equation}
K^2_{*} = - \Delta^{mn} K_m K_n = \frac12 \left(k_v - \sigma^{\prime} \right)^2 \,.
\label{d2}
\end{equation}
Thus,  the phase velocities for the electromagnetic waves with the polarizations along $0x^2$ and $0x^3$ are given, respectively, by the formulas
\begin{equation}
V^{(2)}_{({\rm ph})} = \sqrt{\frac{1-H_{21}}{1+H_{21}}} \,, \quad V^{(3)}_{({\rm ph})} = \sqrt{\frac{1-H_{31}}{1+H_{31}}}\,.
\label{d3}
\end{equation}
Clearly, $V^{(2)}_{({\rm ph})} \neq V^{(3)}_{({\rm ph})}$, when $\beta^{\prime}\neq 0$. Mention should be made that the phase velocity $V^{(2)}_{({\rm ph})}$ does not depend on the plus or minus signs in the solution (\ref{4ch3}) (for $V^{(3)}_{({\rm ph})}$ the situation is similar). For small values of the coupling parameters $\alpha$ and $\gamma$ we obtain
\begin{equation}
V^{(2)}_{({\rm ph})} \simeq 1{-} h_2 = 1 {-} \frac{1}{2\sqrt2}\left[\frac{L^{\prime}}{L}(\alpha{-}\gamma){+} \beta^{\prime}(\alpha{+}\gamma) \right] \,,
\label{d39}
\end{equation}
\begin{equation}
 V^{(3)}_{({\rm ph})} \simeq 1{-} h_4 = 1{-} \frac{1}{2\sqrt2}\left[\frac{L^{\prime}}{L}(\alpha{-}\gamma){-} \beta^{\prime}(\alpha{+}\gamma) \right] \,,
\label{d393}
\end{equation}
so that the difference between the phase velocities of the orthogonally polarized electromagnetic waves
\begin{equation}
V^{(2)}_{({\rm ph})} - V^{(3)}_{({\rm ph})} = h_4- h_2 = -\frac{1}{\sqrt2} \ \beta^{\prime}(\alpha+\gamma)
\label{d38}
\end{equation}
is linear in the factor  $\beta^{\prime}(u)$ and proportional to the sum of the coupling parameters.

\subsubsection{Phase velocities of the electromagnetic waves with the front orthogonal to the pp-wave front}

Based on Eqs. (\ref{t4}) and (\ref{t7}), we obtain for the wave with polarization along $0x^3$
$$
K_i = k \left(\delta_i^{u} \phi^{\prime} + \delta_i^{v} \frac{1}{\sqrt2} - \delta_i^2 \right) \,,
$$
$$
K_m K^m = k^2 \left(\sqrt2 \phi^{\prime} - \frac{e^{-2\beta}}{L^2} \right)\,,
$$
$$
\Omega = \frac{k}{2} \left(1 + \sqrt2 \phi^{\prime} \right) \,,
$$
\begin{equation}
K^2_{*} = \frac{k^2}{2} \left[2\frac{e^{-2\beta}}{L^2} + \left(\phi^{\prime}-\frac{1}{\sqrt2}\right)^2 \right]\,.
\label{d4}
\end{equation}
There are two phase velocities describing the electromagnetic waves with this polarization
$$
V^{(\pm)}_{({\rm ph})} =
$$
\begin{equation}
\frac{ \left[H_{31}-1 \pm \sqrt{1+H_{31}^2 \left(2\frac{e^{-2\beta}}{h_4 L^2}-1 \right)} \ \right]}
{\sqrt{\left[H_{31}{+}1 \mp \sqrt{1{+}H_{31}^2 \left(2\frac{e^{{-}2\beta}}{h_4 L^2}{-}1 \right)} \right]^2 {+} 4 H^2_{31} \frac{e^{-2\beta}}{L^2}}}.
\label{d5}
\end{equation}
These velocities depend on the choice of the sign, in contrast with the longitudinal case. For small values of the coupling parameters $\alpha$ and $\gamma$ Eq. (\ref{d5}) gives
$$
V^{(+)}_{({\rm ph})} \simeq  1 {-} \left[h_4\left(1{+} \frac{e^{-4\beta}}{L^4}\right) {+} 2h_3 \frac{e^{-2\beta}}{L^2} \right] \left[1{+} \frac{e^{-2\beta}}{L^2}\right]^{-2} \,,
$$
\begin{equation}
V^{(-)}_{({\rm ph})} \simeq 1 {-} h_4 \,.
\label{d55}
\end{equation}
For the electromagnetic wave with orthogonal polarization we find that the ordinary wave with $\psi(u) {=}\psi_{(0)}$ propagates with the phase velocity equal to speed of light in vacuum, $V_{({\rm ph})}^{(0)}=1$. For the pair of extraordinary waves we obtain
$$
\tilde{V}^{(\pm)}_{({\rm ph})} =
$$
\begin{equation}
\frac{\left[H_{21}{-}1 \pm \sqrt{\left(1{-}H_{21}^2\right) \left(1 {+} 2h_2 \frac{e^{-2\beta}}{L^2} \right)} \  \right]}{\sqrt{\left[H_{21}{+}1 \mp \sqrt{\left(1{-}H_{21}^2\right) \left(1 {+} 2h_2 \frac{e^{{-}2\beta}}{L^2}\right)}   \right]^2 {+} 4 H^2_{21} \frac{e^{{-}2\beta}}{L^2} }}.
\label{d6}
\end{equation}
When $\alpha \to 0$ and $\gamma \to 0$, this formula gives
$$
\tilde{V}^{(+)}_{({\rm ph})}  \simeq 1 {-} \left[h_2\left(1{+} \frac{e^{-4\beta}}{L^4}\right) {+} 2h_1 \frac{e^{{-}2\beta}}{L^2} \right] \left[1{+} \frac{e^{{-}2\beta}}{L^2}\right]^{{-}2}\,,
$$
\begin{equation}
\tilde{V}^{(-)}_{({\rm ph})}  \simeq 1 - h_2 \,.
\label{d6plus}
\end{equation}
To conclude, one can confirm that electromagnetic waves in the dynamic aether excited by pp-wave modes manifest the effect of birefringence; the proof of this statement is that the corresponding phase velocities depend on polarization and on the propagation direction. Below we consider the birefringence effect in terms of phase shifts for the case when the coupling parameters $\alpha$ and $\beta$ are small enough.

\subsection{On regularity of ordinary and extraordinary waves}

We indicate the electromagnetic wave in the dynamic aether as the regular one, if its phase and amplitude tend to finite values in the limit $\alpha \to 0$, $\gamma \to 0$; the wave can be called anomalous, if at least one of these two quantities tends to infinity.
Using Eq. (\ref{4ch3}) with the plus sign and the minus sign we obtain, respectively,
$$
\sigma^{\prime}_{(+)}  \simeq -  \frac{k_v}{4\sqrt2}\left[\frac{L^{\prime}}{L}(\alpha-\gamma)+ \beta^{\prime}(\alpha+\gamma) \right] \,,
$$
\begin{equation}
 \sigma^{\prime}_{(-)} \simeq -\frac{4\sqrt2 k_v}{\left[\frac{L^{\prime}}{L}(\alpha-\gamma)+ \beta^{\prime}(\alpha+\gamma) \right]}\,.
\label{4ch3plus}
\end{equation}
Using Eq. (\ref{t7}) with the plus sign and minus sign we see that
$$
\phi^{\prime}_{(+)}(u)  \simeq  \frac{e^{-2\beta}}{\sqrt2 L^2}- \frac{1}{2\sqrt2} \left[h_4 \left(1+ \frac{e^{-4\beta}}{L^4} \right)+ h_3 \frac{e^{-2\beta}}{L^2} \right] \,,
$$
\begin{equation}
\psi^{\prime}_{(-)}(u)  \simeq   - \frac{\sqrt2}{h_4} \,. \label{t7plus}
\end{equation}
Similarly, (\ref{t12}) gives
$$
\psi^{\prime}_{(+)}(u)  \simeq \frac{e^{-2\beta}}{\sqrt2 L^2}- \frac{1}{2\sqrt2} \left[h_2\left(1+ \frac{e^{-4\beta}}{L^4} \right)- 2h_1 \frac{e^{-2\beta}}{L^2} \right]  \,,
$$
\begin{equation}
\psi^{\prime}_{(-)}(u)  \simeq  - \frac{\sqrt2}{h_2} \,.
\label{t12plus}
\end{equation}
As for the amplitude functions (\ref{t13}) and (\ref{t14}) with the sign minus, they behave as
\begin{equation}
a^{(-)}_2  \simeq  \frac{\sqrt2  a_v}{h_2} L^2 e^{2\beta}  \,, \quad
a^{(-)}_u \simeq  - a_v  \left(\frac{e^{-2\beta}}{L^2}+ \frac{h_1}{h_2}\right) \,.
\label{t1467}
\end{equation}
This means that the function $a^{(-)}_2$ is large, when the coupling parameters $\alpha$ and $\gamma$ are small; the corresponding double limit at $\alpha \to 0$, $\gamma \to 0$ exists but gives infinity as a result (we deal with the simple pole in standard terminology).  The function $a^{(-)}_u$ remains finite, however, the double limit $\alpha \to 0$, $\gamma \to 0$ does not exist, since the term $\frac{h_1}{h_2}$ depends on the guiding trajectory on the plane $\alpha 0 \gamma$. These are typical symptoms of the existence of an anomaly in the electromagnetic response (see, e.g., \cite{BAlpin,A1,A2,A3,A5} for details).
Thus, for the electromagnetic waves indicated by the minus sign the phase functions are irregular in the limit  $\alpha \to 0$, $\gamma \to 0$. In other words, we are faced with an anomalous behavior of the electromagnetic response on the action of the pp-wave aether mode. A similar anomalous behavior of the electromagnetic response on the action of the gravitational pp-wave we have described in Refs. \cite{BAlpin,A1,A2,A3,A5} by the examples of initially static magnetic and electric fields in media with refraction index $n$ close to one, $n^2 \to 1$. Now we have found examples of anomalous behavior of the phase of the waves propagating in the electromagnetically active dynamic aether.

\subsection{Estimations of the effect magnitude}

In the experiment, the phase of the electromagnetic wave is not a detectable quantity, however, the phase difference of two waves with orthogonal polarizations can be found, e.g., in laser-interferometric systems. For instance, one can try to find the phase differences of the following five wave configurations: {\it first}, when one wave is ordinary, another wave is regular extraordinary; {\it second}, when one wave is ordinary, another wave is irregular extraordinary; {\it third}, when both waves are regular extraordinary; {\it fourth}, when both waves are irregular extraordinary; {\it fifth}, when one wave is regular extraordinary, another wave is irregular extraordinary. In order to estimate the magnitude of the birefringence effect, we consider two examples of the third and fourth types in our classification, calculating the phase differences for the waves with the front parallel to the front of the aether pp-wave mode. According to (\ref{ch5}) they are, respectively,
$$
\Delta W_{({\rm A})} = k_v (\xi_2-\xi_3)=
$$
\begin{equation}
 = k_v \int_0^u  du \left[\frac{\left(1{+}\sqrt{1{-}H_{31}^2} \right)}{H_{31}}-\frac{\left(1{+}\sqrt{1{-}H_{21}^2} \right)}{H_{21}} \right],
\label{plus}
\end{equation}
$$
\Delta W_{({\rm R})} = k_v (\eta_2-\eta_3)=
$$
\begin{equation}
=k_v \int_0^u du \left[\frac{\left(1{-}\sqrt{1{-}H_{31}^2} \right)}{H_{31}}-\frac{\left(1{-}\sqrt{1{-}H_{21}^2} \right)}{H_{21}} \right].
\label{minus}
\end{equation}
When the coupling parameters $\alpha$ and $\gamma$ are small,  we obtain the approximate expressions
$$
\Delta W_{({\rm A})} \simeq 2 k_v \int_0^u du \frac{(H_{21}-H_{31})}{H_{21}H_{31}}  =
$$
\begin{equation}
=8 \sqrt2 k_v  \int_0^u du  \frac{(\alpha + \gamma)\beta^{\prime}(u)}{\left[\left(\frac{L^{\prime}}{L} \right)^2 (\alpha - \gamma)^2- {\beta^{\prime}}^2 (\alpha + \gamma)^2\right]} \,,
\label{plus1}
\end{equation}
\begin{equation}
\Delta W_{({\rm R})} \simeq \frac12 k_v \int_0^u du (H_{31}-H_{21})  = - \frac{(\alpha+\gamma)}{2\sqrt2}k_v  \beta(u) \,.
\label{minus1}
\end{equation}
Clearly, the quantity  $\Delta W_{({\rm R})}$ is small, so the superscript $({\rm R})$ relates to the term {\it regular}, while the phase difference $\Delta W_{({\rm A})}$ is {\it anomalously} large, thus explaining the usage of the  superscript $({\rm A})$. From the physical point of view, the parameter $\Re \equiv \frac{(\alpha+\gamma)}{2\sqrt2}$ is the correlation radius of the electromagnetic interactions attributed to the aether; it has the dimensionality of length. The quantity $\beta$ is in fact the amplitude of the pp-wave aether mode; it can be estimated using the analogy with pure gravitational waves emitted by binary black hole systems, as $|\beta|_{({\rm max})} \propto 10^{-21}$ (the amplitude on the Earth surface).
The quantity $k_v$ for light is of the order $10^{7} {\rm m}^{-1}$, so that the regular phase difference (\ref{minus1}) is of the order $\propto \Re \cdot 10^{-14} {\rm rad}$. Of course, the magnitude of the birefringence effect for the regular case seems to be extremely small, while the corresponding estimate for the anomalous case (see (\ref{plus1})) happens to be extremely optimistic.

\section{Conclusions}

\noindent
1. An electromagnetically active dynamic aether is shown to behave as an anisotropic birefringent medium, when the aether is excited by a pp-wave; ordinary electromagnetic wave and extraordinary electromagnetic waves of two types can propagate in the dynamic aether.

\noindent
2. Electromagnetic waves with orthogonal polarizations, coupled to the pp-wave aether mode, propagate with different phase velocities, thus revealing the birefringence effect induced by the dynamic aether.

\noindent
3. For one of two extraordinary electromagnetic waves, which can be generated in the dynamic aether excited by the pp-wave mode, an anomalous behavior of the phase is predicted; the phase difference calculated for two waves with orthogonal polarizations inherits this anomalous feature.

\vspace{15mm}
\noindent
{\bf Acknowledgments}

\noindent
The work was supported by Russian Science Foundation (Project No. 16-12-10401), and, partially, by the Program of Competitive Growth
of Kazan Federal University.

\end{document}